\documentclass{elsart}
\usepackage{graphicx}
\usepackage[activeacute,english]{babel}
\begin{document}

\begin{frontmatter}
   
\title {
Ratchet, Pawl and Spring Brownian Motor
}
\author{A. Gomez-Marin}
\ead{agomez@ecm.ub.es}
\author{and}
\author{J. M. Sancho}
\ead{jmsancho@ecm.ub.es}

\address{Departament d'Estructura i Constituents de la Materia, Facultat de
  Fisica, Universitat de Barcelona, Diagonal 647, 08028 Barcelona, Spain} 


\begin{abstract} 
We present a model for a thermal Brownian  motor based on Feynman's famous ratchet and pawl
device. Its main feature is that the ratchet and the pawl are in different 
thermal baths and connected by an harmonic spring.
We simulate its dynamics, explore its main features  and also derive an approximate 
analytical
solution for the mean velocity as a function of the external torque applied
and the temperatures of the baths. Such theoretical predictions and the
results from  
numerical simulations agree within the ranges of the approximations
performed. 
\end{abstract}


\end{frontmatter}



\section{Introduction}

The engines which aim to get useful work by rectifying thermal
fluctuations are called Brownian motors (BM).
During the last years a lot of effort has been invested  to
study the underlying mechanism of such engines which has been called as the
ratchet 
effect \cite{reiBM}. This is a 
mechanism which consists in breaking the spatial and temporal inversion 
symmetry of the
system so that directed transport emerges, being thermal fluctuations
the very  
relevant input. In fact, the paradigmatic device of such speculations is
Feynman's famous ratchet and pawl machine \cite{feyn}.

\begin{figure}
\begin{center}
 \includegraphics[ width=0.55\textwidth]{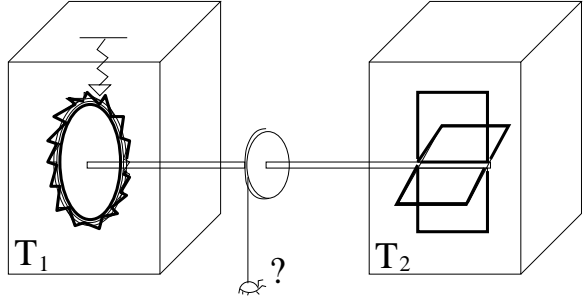}
   \caption{Feynman's original ratchet and pawl machine.}
   \label{frp}
\end{center}
\end{figure}

In 1963 R. P. Feynman introduced \cite{feyn} a microscopic 
device  (the ratchet and pawl machine) that  can operate between two thermal 
baths extracting some mechanical work. 
The hotter bath contains an axle with vanes in it. The
bombardments of gas
molecules on the vane make the axle rotate with random symmetric fluctuations.
At the other end side of
the axle, as shown in Fig. (\ref{frp}), there is  a second box with an
asymmetric toothed wheel which in principle can turn only one way due to a
coupling with a  pawl (the stopping mechanism).  

At first glance one could think that it seems quite likely
that the wheel will spin round one way and lift a weight even
when both gases are at the same temperature and thus violating the Second Law. 
However,
a closer look at the pawl reveals that it bounces and  so the wheel will rotate
randomly in any
direction, doing a lot of jiggling but with no net turning. Thus, the machine
cannot extract work from two baths at the same temperature.
When the temperature of the vanes is
higher than the temperature of the wheel, Feynman concludes that
some work is performed with Carnot's
efficiency when the machine is lifting the weight very slowly.
This is indeed a very optimistic
result which has been revised,
many years later, in Refs. \cite{parr,magfrp}.
Such particular device has been analyzed and it has been
suggested that this engine is very far from Carnot efficiency
\cite{parr} due to the fact that this type  of device has strong heat
losses. 
A more refined analysis and modeling of Feynman's motor has been
presented in  
several references \cite{magfrp,SBM,seki}, with the conclusion that its 
efficiency is very poor. 
The main problem comes from the mechanical coupling of the pawl mechanism, 
which is not very efficient.

Moreover, the main idea underlined in the the ratchet and pawl mechanism can
be  implemented in different ways. 
In this work we propose a model through an equation for  a 
dynamical classical variable (position 
or angle) moving in a
periodic and asymmetric potential (a ratchet potential) in a bath at
temperature $T_{1}$ coupled with 
another degree of freedom immersed in a different bath at temperature $T_{2}$.
The coupling mechanism is a harmonic spring.  
In this case
the two thermal baths are clearly separated and the performance of this engine,
as a function of the different parameters, can be studied.
In the following section, we introduce the model. Once the model is
explained and justified, we present in  
Section 3 
results by numerical simulations of the stochastic equations of motion.
In Section 4 an analytical approach is presented and in Section 5 we show
the agreement of the theoretical expressions found with the 
data from numerical simulations. Finally we 
end with some conclusions and comments.

\section{The Ratchet, Pawl and Spring Motor Brownian Motor (RPSBM)}

\begin{figure} [t]
\begin{center}
  \includegraphics[ width=0.65\textwidth]{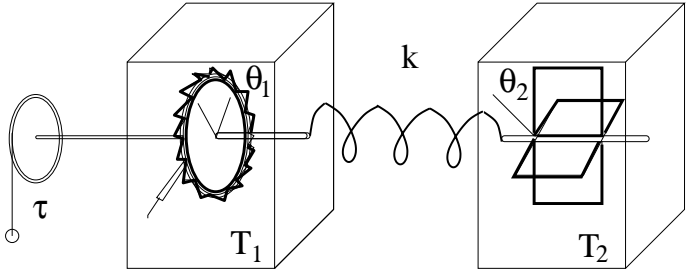}
  \caption{Ratchet, Pawl and Spring Brownian Motor.}
  \label{fig1}
\end{center}
\end{figure}

First of all, let's describe our proposal for a Brownian motor as it is shown 
in  Fig. (\ref{fig1}). It consists of
two boxes at different constant temperatures.
The left box contains a ratchet and a pawl that act like a 
mechanical
rectifier device. From this box an axel-wheel device is
connected allowing it to lift a hanging object in order to study how much
work the device can perform. The second box is at a higher temperature than
the first one and has a little windmill that is used to pick up energy from
the thermal bath (through the collisions of the particles of the bath with the
vanes) and to transfer it to the ratchet and pawl system through the spring.  
The main simplification of our model is that the dented wheel and the pawl 
mechanisms are substituted by a ratchet potential and a harmonic spring.
Then the Langevin equations in the over-damped limit are straightforward to 
write,
\begin{equation}\label{eq1}
\lambda_{1}\frac{d\theta_{1}}{dt}=-\frac{\partial V_{R}(\theta_{1})}{\partial
    \theta_{1}}-k(\theta_{1}-\theta_{2})-\tau+\xi_{1}(t), 
\end{equation}
\begin{equation}\label{eq2}
\lambda_{2}\frac{d\theta_{2}}{dt}=k(\theta_{1}-\theta_{2})+\xi_{2}(t),
\end{equation}
with thermal noise satisfying the fluctuation--dissipation relation,
\begin{equation}
\langle \xi_{i}(t)\xi_{j}(t') \rangle
=2k_{B}T_{i}\lambda_{i} \delta_{ij}\delta(t-t'). 
\end{equation}
Note that we are describing the evolution in time of the angular  positions 
$\theta_{1}$ and $\theta_{2}$ of
the ratchet and the windmill respectively. We see in equation (\ref{eq1}) a
force term that comes from a periodic and asymmetric potential $V_{R}$ that
models the shape of the sawtooth wheel. There's also an external torque
$\tau$ that can be different than zero to account for the work done.
 Two independent white noises $\xi_{1}$ and $\xi_{2}$
represent thermal fluctuations and, finally, there's an interaction
between the two baths through a harmonic spring of constant $k$. A linear
coupling between both degrees of freedom also appears in
Ref. \cite{cilla}.

\begin{figure}
\begin{center}
\includegraphics[angle=270, width=0.5\textwidth]{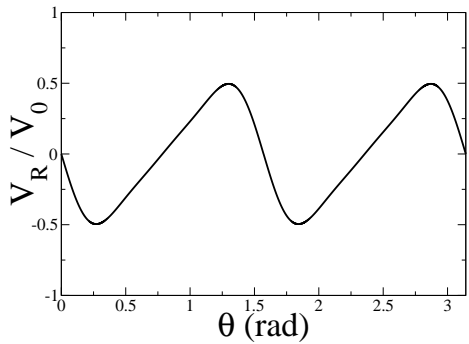}
\caption{Shape of the ratchet potential $V_{R}$ for $d=4$.}
\label{rat}
\end{center}
\end{figure}

The ratchet potential $V_{R}(\theta)$ is given by,
\begin{equation} \label{pot}
V_{R}(\theta) = -\frac{V_{0}}{2.23}[\sin(d
\theta)+0.275\sin(2d\theta)+0.0533\sin(3d\theta)], 
\end{equation}
where $V_{0}$ controls the height of the
potential,  $d$ is the number of teeth per turn, and
the asymmetry of the potential is controlled  by changing the numerical
coefficients that multiply the
sinus functions.
In Fig. (\ref{rat}) we see the explicit form of the ratchet potential 
$V_{R}(\theta)$ used 
in our study.

This problem has a set of parameters which can be simplified if a new 
time dimensionless scale is defined, $t=\frac{\lambda_{1}}{Vo} s$. 
Then the equations (\ref{eq1}) and (\ref{eq2}) are transformed and all  
the relevant parameters: $\widetilde{T_{1}}=\frac{k_{B}T_{1}}{Vo}$,
$\widetilde{T_{2}}=\frac{K_{B}T_{2}}{Vo}$, $\widetilde{k}=\frac{k}{Vo}$,
$\widetilde{\lambda}=\frac{\lambda_{1}}{\lambda_{2}}$ and
$\widetilde{\tau}=\frac{\tau}{Vo}$, are now  dimensionless.
In this way we can concentrate on the main parameters
that control the dynamics of the system. For instance, we see that $V_{0}$ controls
the energy scale, with respect to the thermal energy of the baths.

\section{Numerical results and optimal regime of the motor}

\begin{figure} 
\begin{center}
       \includegraphics[angle=270, width=0.6\textwidth]{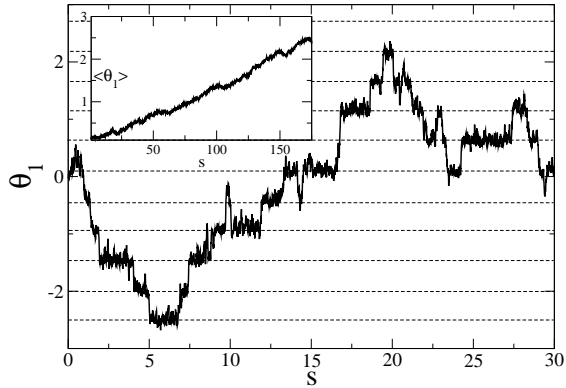}
   \caption{Typical evolution of the angle position $\theta_{1}$ as a function
     of the dimensionless time $s$ directly obtained from the
     simulations. The parameters are $V_{0}=3k_{B}T_{1}$, $k=100k_{B}T_{1}$,
      $d=12$ and $T_{2}=2T_{1}$. Plot of a single trajectory
     (for $\tau=0$) in which one can see the discrete jumps of the dynamics
     from one valley to another of the saw-tooth potential. Inset: 
     Average of $\theta_{1}(s)$ over many particles for
     $\tau=0.01k_{B}T_{1}$.}  
   \label{runs}
\end{center}
 \end{figure}

In this section we present the results by numerically simulating
Eqs. (\ref{eq1}) and (\ref{eq2}). We have used a second 
order algorithm (Heun), which is a generalization of the Runge-Kutta
algorithm for stochastic systems. Since the process is intrinsically
nondeterministic, it is convenient to do statistics in order to minimize
fluctuations in the output data.
For that reason, we have averaged every step of the integration over 500
different particles. 
In figure (\ref{runs}) we show an example of a single trajectory in time and
the mean $\langle \theta_{1}(s) \rangle$.

We must first explore  the role played by every parameter of the model 
in order to get in scale and tune them cleverly to obtain the most efficient 
working regime. The height of the potential $V_{0}$ is a very relevant
parameter since it  determines the energy scale of the system. 
We are interested in an energy barrier
higher than the energies of the two thermal baths but not much
more. Otherwise, the 
engine wouldn't work because thermal fluctuations would not be able to cause a 
jump over such a high barrier.
There is an optimal value that is found to be around $V_{0}=3K_{B}T_{1}$. For
smaller values, the height is too small and thus fluctuations can jump it over
to both sides, giving a zero velocity on average. For large $V_{0}$, hardly
ever is the system able to perform a jump to the next valley. This is what is 
expected from intuition and also what is found from numerical simulations. 

Regarding temperatures, we must choose $T_{2}$ higher than $T_{1}$ but not
too much because large randomness \cite{klu} blurs and destroys directional
motion. In a following section we will analyze in detail the behavior of the
mean speed as a function of the ratio of temperatures. In fact, there is
an optimal ratio which is around $T_{2}\simeq 4T_{1}$.

We have also checked that the velocity falls to
zero for two limits of the coupling constant: $k\to 0$ and $k\to \infty$.
This is in perfect accordance with the following thermodynamical
arguments. For $k$ going
to zero, the systems are un-coupled and therefore 
we have a single bath at a constant temperature. In this
situation the wheel will have no mean
velocity. For the opposite limit, this is, large values of the coupling
constant, we  
use the argument discussed in Ref. \cite{parr}. A very high $k$ is 
equivalent to
having both systems joined by a rigid axel vane. Then, the system is in
contact with two thermal baths at different temperatures. One can prove that
the system feels a single averaged temperature and thus it will not run at all
on average either. 

There seems to be an optimal number of teeth $d$ as well. For
low values of 
$d$ the wheel isn't a proper sawtooth and  
the asymmetry of the potential is not felt. For $d$ around $d=24$. we find
that the wheel spins quicker, while for $d$ 
greater the velocity decays again. Notice that the force term is
proportional to the number of teeth. This means than the higher $d$
is, the steepest the potential becomes.

Using three other different models
of the  ratchet potential, which are built  similarly
but modifying the coefficients multiplying the sinus functions, we see that the
more asymmetrical the model is, the faster the system runs. 
Obviously for a symmetric potential no rectification of the
fluctuations is possible for any value of the rest of parameters.

When examining equations (\ref{eq1}) and (\ref{eq2}) in
dimensionless variables, one finds that the noises, in dimensionless
variables too, should be written as
\begin{equation}
\langle \widetilde{\xi_{1}}(s)\widetilde{\xi_{1}}(s') \rangle
=2\widetilde{T_{1}}\delta(s-s'), 
\end{equation}
\begin{equation}
\langle \widetilde{\xi_{2}}(s)\widetilde{\xi_{2}}(s') \rangle
=2\widetilde{\lambda}\widetilde{T_{2}}\delta(s-s'), 
\end{equation}
\begin{equation}
\langle \widetilde{\xi_{1}}(s)\widetilde{\xi_{2}}(s') \rangle=0.
\end{equation}
Notice that the effective intensity of the second bath is the thermal energy
of the bath in terms of $V_{0}$ multiplied by the fraction of the friction
coefficients.

Finally, when setting $\lambda_{1}=\lambda_{2}$, $T_{2}=2T_{1}$, 
$V_{0}=3 K_{B}T_{1}$, $\tau=0$, $k=80 k_{B}T_{1}$ and $d=24$,
we are in an optimal regime in which the an angular velocity is nearly the
greatest possible: $v \doteq \langle \dot{\theta_{1}} \rangle \simeq 0.021 s^{-1}$.

Apart from the quantitative numerical value itself, we can make two main
conclusions. The first one is that the engine does work. We have made some
simulations in 
which we use a torque $\tau$ different than zero. For instance, for a
small torque $\tau=0.1k_{B}T_{1}$ we see that the systems still runs and 
lifts the external weight, thus performing useful work. 
The second conclusion
is that this motor is very inefficient. Energetics are not
worth to be calculated in detail because just by comparison to the model
in \cite{SBM} we can see that the maximum velocity achieved by the RPSBM
is at least five times smaller than the speeds found in
Ref. \cite{SBM} for the SBM. Therefore the efficiency of the present model
is even much smaller.

\section{Analytical study of the RPSBM}

Our purpose 
now is to make an analytical study of our motor. Any kind of
exact calculation in such systems seems impossible and, therefore, some  
approximations have to be assumed.

Consider the equations that define our Brownian motor written 
in the form

\begin{equation}
\dot{\theta_{1}}=f(\theta_{1})- k(\theta_{1}-\theta_{2})-\tau 
+\xi_{1}(t),
\end{equation}
\begin{equation}
\dot{\theta_{2}}=  k(\theta_{1}-\theta_{2}) +\xi_{2}(t),
\end{equation}
where $f(\theta_1)$ is the force exerted by the ratchet potential:
$f(\theta_1)=-V_{R}'(\theta_1)$. Note that we have set
the friction coefficients to one for simplicity without losing any generality
in the calculus.
Let us define now the changes of variables 
\begin{equation}
x=\frac{\theta_{1}+\theta_{2}}{2}, \qquad
y=\frac{\theta_{1}-\theta_{2}}{2}.
\end{equation}
The relevant variable $x$ describes the evolution of the center of mass
and the "irrelevant" variable $y$  describes the relative motion of   
the two-particle system.
The equations of motion in these new variables are 
\begin{equation}
\dot{x}=\frac{f(x+y)}{2}-\frac{\tau}{2}+\eta_{1}(t),
\end{equation}
\begin{equation}
\dot{y}=\frac{f(x+y)}{2}-\frac{\tau}{2}-2ky+\eta_{2}(t),
\end{equation}
where a redefinition of the noises has been introduced as
\begin{equation}
\eta_{1}(t)=\frac{\xi_{1}(t)+\xi_{2}(t)}{2}, \qquad
\eta_{2}(t)=\frac{\xi_{1}(t)-\xi_{2}(t)}{2}.
\end{equation}

Let's make our first approximation. Using the fact that $y$ is very
small, we can make a Taylor expansion of $f(x+y)$ up to first order in $y$,
obtaining the pair of equations
\begin{equation}
\dot{x}=\frac{f(x)+yf'(x)-\tau}{2}+\eta_{1}(t),
\label{xeq}
\end{equation}
\begin{equation} 
\label{yeq}
\dot{y}=\frac{f(x)-\tau+ y(f'(x)-4k)}{2}+\eta_{2}(t).
\end{equation}

Since numerical simulations indicate that the variable $y$ has a faster
dynamics we will  
eliminate  it
adiabatically. This means we set $\dot{y} \doteq 0$. Then
Eq. (\ref{yeq}) reduces to
\begin{equation}
y = \frac{1}{4k-f'(x)} [f(x)-\tau+2\eta_{2}(t)].
\end{equation}
 One can see that the term $4k$ is much bigger than $f'(x)$ when
 $k \simeq 100k_{B}T_{1}$ 
and $d$ is small ($d\simeq4$). Then, within this range of parameters we can
 keep only $4k$ in the denominator. 
Substituting now the last expression for $y$ in equation (\ref{xeq}) we end up
with a Langevin equation with a new force $H(x)$ and two
multiplicative noises $g_{1}(x)$ and $g_{2}(x)$, 
\begin{equation} \label{eqx2}
\dot{x}= H(x)+g_{1}(x)\xi_{1}(t)+g_{2}(x)\xi_{2}(t),
\end{equation}
where,
\begin{equation} \label{Heq}
H(x) =\frac{1}{2}(f(x)-\tau)(1+\frac{f'(x)}{4k}),
\end{equation}
\begin{equation}
g_{1}(x)=\frac{1}{2} \left( 1+\frac{f'(x)}{4k} \right), \qquad 
g_{2}(x)=\frac{1}{2} \left( 1 -\frac{f'(x)}{4k} \right).
\end{equation}

The
Fokker-Planck equation associated to Eq. (\ref{eqx2}) is
\begin{equation}
\partial_{t}P(x,t)=-\partial_{x}J(x,t),
\end{equation}
where
\begin{eqnarray}
& J(x,t)=H(x)P(x,t) -k_{B}T_{1} [g_{1}(x)\partial_{x}g_{1}(x)P(x,t)]  &   
\nonumber \\
&-k_{B}T_{2}[g_{2}(x)\partial_{x}g_{2}(x)P(x,t)]. &  
\end{eqnarray}
After some manipulations
with partial derivatives, the probability current $J(x,t)$ can be rewritten as
\begin{equation} \label{flux}
J(x,t)=H(x)P(x,t)-[g_{eff}(x)\partial_{x}g_{eff}(x)P(x,t)], 
\end{equation}
with
\begin{equation} \label{geff}
g_{eff}(x)= \sqrt{k_{B}T_{1}g_{1}^{2}(x)+k_{B}T_{2}g_{2}^{2}(x) }.
\end{equation}

Since we are interested in the steady state, we have to solve now equation
(\ref{flux}) for  
a constant probability current $J$. The first step is to reduce this equation
to a Bernoulli form which can be formally integrated. By imposing periodic 
boundary conditions, $P(x)=P(x+L)$, where $L=\frac{2\pi}{d}$, we find that
\begin{equation} 
\label{JP}
P_{0}(1-e^{\beta(L)})= J\int_{0}^{L} dx \frac{e^{-\beta(x)}}{g_{eff}(x)},
\end{equation}
where $P_{0}$ is a constant that can be found from the normalization
condition 
$\int_{0}^{L}P(x)dx=1$, and $\beta$ is a relevant function whose
expression is 
\begin{equation} 
\label{beta1}
\beta(x)=\int_{0}^{x}dx'\frac{-H(x')}{g_{eff}^{2}(x') }.
\end{equation}

The mean velocity $v \doteq \langle \dot{x} \rangle$ is just the
integral of the probability current over the spatial period $L$. Since 
$J$ is constant, we have from Eq. (\ref{JP}) that the solution for
$v$ is \cite{rei,risk,lutz,buti1,buti2} 
\begin{equation}
v= \mathcal{N} (1-e^{\beta}),
\label{vtheo}
\end{equation}
where $\mathcal{N}$ is a constant that is found by using the normalization
condition for the probability $P(x)$. We will see that this 
constant depends very smoothly on the control parameters of our model.
Notice that the mean velocity $v \simeq \langle \dot{\theta_{1}} \rangle $
because $ \langle \dot{\theta_{1}} \rangle \simeq \langle
\dot{\theta_{2}} \rangle $.

At this point, what is left to do is to find the $\beta$ integral as a
function of the parameters we are interested in. From now on we will assume
implicitly that the energy is in $k_{B}T_{1}$ units. 
To simplify the calculation of the integral, we make an expansion of the
denominator in equation (\ref{beta1}) in powers of
$\frac{1}{k}$. Since the value of $k$ that we will consider is the one that
makes the motor run faster ($k\simeq 100$), one can safely suppose that the
terms of the order $(\frac{1}{k})^{2}$ and so on will not notably contribute
to the integral. Remember that $d$ and $V_{0}$ are kept small. 
The relevant quantity $\beta$ can then be written as
\begin{equation} 
\beta=\int_{0}^{L} dx \frac{\frac{1}{2}(-f(x)+\tau)(1+\frac{f'(x)}{4k})} {
    \frac{1}{4}[  1+\frac{T_{2}}{T_{1}}
    +2(1-\frac{T_{2}}{T_{1}})\frac{f'(x)}{4k} ]}. 
\end{equation}
Let's note that the terms $-f(x)+\tau$ in the numerator are very small when
integrated. Therefore we can neglect the much smaller correction
$f'(x)/4k$ of the numerator. However, one cannot do this approximation for the
same term in 
the denominator because it is the responsible from the net motion of the
Brownian motor. Such term 
contains two essential physical features. The first one is that it accounts
for the multiplicative noise and thus, without it, thermal sources are unable to make
the motor move. The second one is that it avoids the
violation of the Second Law, i.e. when $T_{2}=T_{1}$ it cancels and no average
velocity is predicted. Then, the final and simplest expression for $\beta$
is  
\begin{equation} 
\label{beta}
\beta=2\int_{0}^{L} \frac{-f(x)+\tau} {
      1+\frac{T_{2}}{T_{1}}+2(1-\frac{T_{2}}{T_{1}})\frac{f'(x)}{4k} }dx.
\end{equation}

\section{Simulations versus theory}

\begin{figure}
\begin{center}
 \includegraphics[ width=0.45\textwidth, angle=270]{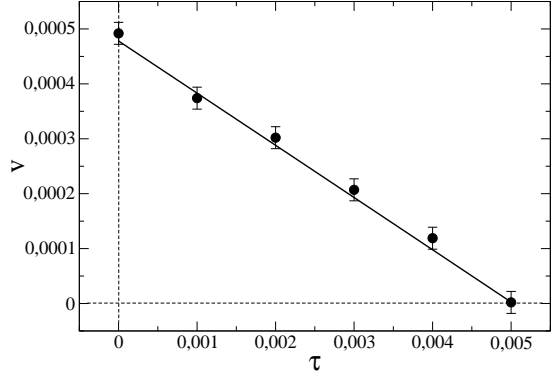}
   \caption{Comparison between numerical simulations (dots) and theoretical
     prediction (line) from (\ref{vteor}) of the dependence of the mean
     velocity $v$ as a function of the external torque applied $\tau$. The
     parameters chosen are 
     $k=100$, $d=4$, $V_{0}=3$ and $T_{2}=2T_{1}$.}
   \label{vtau}
\end{center}
\end{figure}
 
Now that we have derived an analytical expression for the
mean velocity of the motor, let's compare its predictions to the numerical simulations when
exploring two relevant parameters: the external torque $\tau$ and the
temperature difference ratio $T_{2}/T_{1}$.

The first comparison is plotted in figure
(\ref{vtau}), in which we show the mean velocity as a function of the external
torque. It is very remarkable that the stall torque (the torque at which the
motor stops) is perfectly predicted by the expression found. What is more, the
functional behavior (the linear dependence with negative slope) is clearly
reproduced. To determine the scale we adjust $\mathcal{N}$ so that the
analytical prediction fits the simulations at $\tau=0$, at the same time that
we confirm that such constant $\mathcal{N}$ does not depend very
much on $\tau$. 

Since $\beta$ is found to be very small, one can write the
following expression for the mean velocity,
\begin{equation} \label{eqfit}
v=\mathcal{N}(1-e^{\beta}) \simeq
\mathcal{N}(-\beta)=a-\tau \; b,
\end{equation}
where
\begin{equation} \label{aterm}
a=2\mathcal{N} \int_{0}^{L} \frac{f(x)} {
      1+\frac{T_{2}}{T_{1}}+2(1-\frac{T_{2}}{T_{1}})\frac{f'(x)}{4k} }dx,
\end{equation}
\begin{equation} \label{bterm}
b=2\mathcal{N} \int_{0}^{L} \frac{1} {
      1+\frac{T_{2}}{T_{1}}+2(1-\frac{T_{2}}{T_{1}})\frac{f'(x)}{4k} }dx.
\end{equation}
This last expression for the parameter $b$ can be easily simplified by
expanding its denominator and stopping at the first term, finding 
\begin{equation}
b \simeq
\frac{2\pi}{d}\frac{2\mathcal{N}}{1+\frac{T_{2}}{T_{1}}}.  
\end{equation}
The term $a$ is a little more complicated but
it can be computed numerically. Finally we obtain  
\begin{equation} \label{vteor}
v \simeq 0.00049-0.097\tau,
\end{equation}
in units of $s^{-1}$. This result is plotted in Fig. (\ref{vtau}).

\begin{figure} 
\begin{center}
 \includegraphics[ width=0.45\textwidth, angle=270]{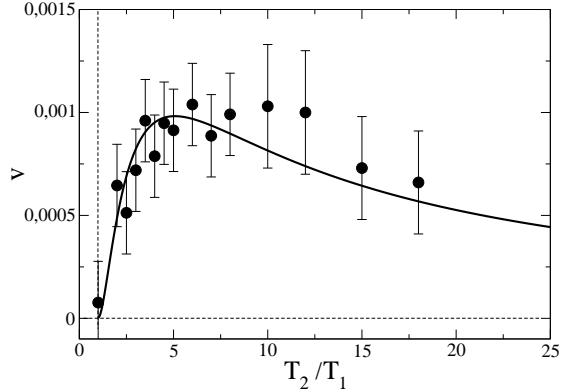}
   \caption{Mean velocity $v$ as a function of the ratio of temperatures
     $T_{2}/T_{1}$. The analytical prediction (line) fits the numerical
     results (dots) quite well allowing for the fact that there's some
     uncertainty in the data obtained from the simulations. In this case
     $\tau=0$ and the other control parameters are unchanged.}
   \label{vt2}
\end{center}
\end{figure}

Our second result is shown in figure (\ref{vt2}), in which the
prediction for the mean velocity is $v=a$, as one finds when setting $\tau=0$
in Eq. (\ref{eqfit}). Despite 
data precision difficulties in the simulations, we can say 
that theory and simulations agree quantitatively and  
also reproduce three important qualitative facts. The first is that at
$T_{1}=T_{2}$ the mean velocity is zero, as we explained before. The second is that
there is a maximum of the speed (an optimal value for the ratio of
temperatures). Thirdly, we observe a slow decay of the velocity to 
zero as $T_{2}$ is increased.
We underline as well that the constant $\mathcal{N}$ used now is the
same that we found for the $v(\tau)$ plot in Fig. (\ref{vtau}). This again 
confirms our hypothesis that $\mathcal{N}$ depends very smoothly on the parameters we are
exploring ($\tau$ and $T_{2}/T_{1}$).
When expanding the denominator of equation (\ref{aterm}), we find that
the terms appear as a function of the ratio $T_{2}/T_{1}$ and its powers.
Therefore, we can roughly see that
this more complex dependence of $v(T_{2}/T_{1})$ comes from such
terms. However, one must keep in mind that it is not a good idea to try to
find exactly which these terms are since our very first approximations killed
terms of the same order in $k$.
Nevertheless, we can say that our analytical predictions reproduce very well
the numerical results. They capture the main qualitative behavior and also fit
the quantitative data.

\section{Conclusions}

We have presented and studied a model for a thermal Brownian motor inspired on
Feynman's ratchet and pawl \cite{feyn}. 
After explaining Feynman's idea we have introduced and justified the RPSBM
model and its main features have been explored numerically, finding the
optimal regime of the motor. The results of the simulations are
consistent with fundamental physical arguments that must always hold.
We have performed an analytical calculation based on \cite{SBM} with
appropriate approximations  
to get an expression for the mean velocity in terms of the relevant
parameters of the model. We have analyzed its dependence on the
external torque $\tau$ 
and the ratio of temperatures $T_{2}/T_{1}$. Such formal predictions fit very
well the data from the numerical simulations. 

Regardless of the particular properties of these kind of heat
engines, they are anyhow unrealistic models for molecular motors \cite{Oster} 
since it is
known that such biological systems are mono-thermal and convert chemical energy into work,
without the intermediate state of burning fuel. Consequently, one cannot think
of these models as realistic ones for biological molecular motors.
Moreover, the mechanical coupling mechanism between both baths acts as a very
good heat conductor even in situations of very small mean velocity. Therefore,
the efficiency is  
only a small fraction of that of Carnot. What is more, it is much smaller than
the efficiency found in similar models \cite{SBM}, as one can see from the
values of the velocity $v$.
Such very low efficiency is a general feature of
Brownian motors due to the fact that, in order to rectify thermal
fluctuations, these systems must be tightly connected and then a lot of heat
is interchanged. 

This research was supported by the Ministerio de
Educaci\'on y Ciencia (Spain) under project BFM2003-07850-C03-01
and the Generalitat de Catalunya (Spain) under grant 2005FI 00194.

\end{document}